\documentclass[12pt,epsfig]{article}
\usepackage{amssymb}
\usepackage{epsf}
\usepackage{lscape}
\usepackage[cp1251]{inputenc}
\usepackage[T2A]{fontenc}
\usepackage[english]{babel}
\usepackage[dvips]{graphicx}
\usepackage{amsmath}
\begin{document}

\begin{center} 
{\bf \uppercase{Quantization of the kinetic energy of a deformed nucleus in curvilinear coordinates}}
 \end{center}

\begin{center}
M. S. Nadirbekov$^1$, O. A. Bozarov$^1$ and  N. Minkov$^2$
\end{center}
\begin{center}
\emph{$^1$Institute of Nuclear Physics, 100214, Tashkent, Uzbekistan}
\end{center}
\begin{center}
\emph{$^2$Institute for Nuclear Research and Nuclear Energy, Bulgarian Academy of Sciences, Tzarigrad Road 72, BG-1784 Sofia, Bulgaria}
\end{center}
{\footnotesize 
The quantization of the kinetic energy of a deformed nucleus in curvilinear coordinates in the case of octupole oscillations of its surface firstly has been carried out.
The obtained form of the Hamiltonian differs from the previously obtained Hamiltonian for quadrupole oscillations only by factors in front of the derivatives $\partial/\partial\gamma$ and $\partial/\partial\eta$.  An explicit form of the kinetic energy of the Hamiltonian of even-even nuclei with free and effective triaxiality, as well as for axially symmetric even-even nuclei, is given.  }

\begin{center}
{\bf  \uppercase{I. Introduction }}
\end{center}
The main excitation modes of nuclei correspond to collective types of motion, such as surface vibrations and elastic vibrations. The observed nuclear levels with certain vibrational modes provide a direct explanation for the rapid increase in the level density with increasing excitation of the nucleus \cite{bohr}. In it, the core surface is described by five expansion parameters corresponding to second-order spherical harmonics for quadrupole deformations. And in the work \cite{davyd} the total energy of quadrupole and octupole deformations is obtained, where the core surface is described by seven expansion parameters corresponding to spherical harmonics of the 3rd order.

It is known that, along with Cartesian rectangular coordinate systems, one can use curvilinear coordinate systems \cite{tikhon}. In certain cases, curvilinear coordinates are more convenient than Cartesian ones. If the work in \cite{greiner} using the Pauli procedure \cite{pauli} describes in detail the quantization of the kinetic energy of a deformed nucleus in curvilinear coordinates for quadrupole oscillations of the surface of the nucleus. The quantization of the classical kinetic energy associated with  the octupole oscillations of the surface of the nucleus has not been performed.

The aim of the present work is to quantize the classical kinetic energy associated with the octupole oscillations in curvilinear coordinates. 

\begin{center} 
{\bf \uppercase{II. Quantization of the kinetic energy of a deformed nucleus in curvilinear coordinates}}
 \end{center}
We take the classical kinetic energy in the form  \cite{greiner}
\begin{equation}
T=\frac{1}{2}B\sum_\mu\dot{\alpha}^*_{2\mu}\dot{\alpha}_{2\mu}.
\end{equation}
Expressing it in terms of Euler angles and internal variables, we obtain
\begin{equation}
T=\sum_k\frac{{M'}_k^2}{2J_k(\alpha_\nu)}+\frac{1}{2}B_3(\dot{a}^2_{30}+2\dot{a}^2_{32})
\label{k1}
\end{equation}
where
\begin{equation}
a_{30}=\beta_3\cos\eta,\hspace{0.3cm}a_{32}=\beta_3\frac{\sin\eta}{\sqrt{2}},
\label{a20}
\end{equation}
 $J_k$--moment of inertia: 
\begin{eqnarray}
\label{j12}
J^{(3)}_1=B_3(6a^2_{30}+2\sqrt{30}a_{30}a_{32}+8a_{32}^2)=4B_3\beta^2_3(\sin^2\eta+
\frac{\sqrt{15}}{2}\sin\eta\cos\eta+\frac{3}{2}\cos^2\eta),\\
J^{(3)}_2=B_3(6a^2_{30}-2\sqrt{30}a_{30}a_{32}+8a_{32}^2)=4B_3\beta^2_3(\sin^2\eta-
\frac{\sqrt{15}}{2}\sin\eta\cos\eta+\frac{3}{2}\cos^2\eta),\\
J^{(3)}_3=8B_3a_{32}^2=4B_3\beta^2_3\sin^2\eta.
\label{j13}
\end{eqnarray}

The expression for the kinetic energy (\ref{k1}) is classical, so it is necessary to quantize the classical Hamilton function in curvilinear coordinates.

The quantum expression for the kinetic energy has the form
\begin{equation}
T=-\frac{\hbar^2}{2}\sum_{\mu\nu}\frac{1}{G^{1/2}}\frac{\partial}{\partial q_\mu}G^{1/2}g^{-1}_{\mu\nu}\frac{\partial}{\partial q_\nu},
\label{k2}
\end{equation}
where $G$--matrix determinant $q_{\mu\nu}$  and $q^{-1}_{\mu\nu}$ denotes the matrix inverse of the matrix $q_{\mu\nu}$. The volume element is 
\begin{equation}
d\tau=|\sqrt{G}|dq_1.....dq_N.
\label{vol}
\end{equation}
To quantize the expression (\ref{k1}), we use the formula (\ref{k2}) We rewrite (\ref{k1}) in the form
\begin{equation}
T=\frac{1}{2}\sum_{k}J^{(3)}_k(a_{30},a_{32})\left(\sum_iV_{kl}(\theta_1,\theta_2,\theta_3)\frac{d\theta_i}{dt}\right)^2+\frac{1}{2}B_3(\dot{a}^2_{30}+\dot{a}^2_{32}),
\label{k3}
\end{equation} 
where  \cite{greiner}
\begin{equation}
V_{k,l}(\theta_1,\theta_2,\theta_3)= \left(
\begin{array}{ccc}
-\sin\theta_2\cos\theta_3 &\sin\theta_3&0\\
\sin\theta_2\cos\theta_3 &\cos\theta_3&0\\
\cos\theta_2 &0&1.
\end{array}
\right)
\end{equation}
Hereof
$$
g_{11}=B_3,\hspace{0.3cm}g_{22}=2B_3,
$$
$$
g_{1k}=g_{k1},\hspace{0.3cm} \text{дл } \hspace{0.3cm}k\neq 1,
$$
$$
g_{2k}=g_{k2}, \hspace{0.3cm}\text{дл } \hspace{0.3cm}k\neq 2.
$$
\begin{equation}
g_{\mu\nu}=\sum_k J_kV_{k\mu}(\theta_i)V_{k\nu}(\theta_i), \hspace{0.5cm}\mu, \nu\geq 3.
\end{equation}
Explicit expression for this matrix
\begin{equation}
g_{\mu\nu} = \left(
\begin{array}{ccccc}
B_3 &0&0&0&0\\
0 &2B_3&0&0&0\\
0 &0&(J_1\cos^2\theta_3+J_2\sin^2\theta_3)\sin^2\theta_2+J_3\cos^2\theta_2&
(J_2-J_1)\sin\theta_2\cos\theta_3\sin\theta_3&J_3\cos\theta_2\\
0 &0&(J_2-J_1)\sin\theta_2\cos\theta_3\sin\theta_3&
J_1\sin^2\theta_3+J_2\cos^2\theta_3&0\\
0 &0&J_3\cos\theta_2&0&J_3.
\end{array}
\right)
\label{mat}
\end{equation}
Easy to calculate matrix determinant
\begin{equation}
G=det g_{\mu\nu}=2B^2_3\sin^2\theta_2J_1J_2J_3=64B^5_3\sin^2\theta_2 a_{32}^2\left[\left(3a^2_{30}+4a^2_{32}\right)^2-30a^2_{30}
a^2_{32}\right].
\label{det}
\end{equation}

It can be seen from the expression (\ref{k2}) that the quantum expression for the kinetic energy is determined by the inverse matrix $g^{-1}_{\mu\nu}$, and its definition requires the concepts of a transposed matrix, a matrix minor, and an algebraic complement of a matrix element .

The matrix inverse to the matrix ${\displaystyle A}$ can be represented as:
\begin{equation}
A^{-1}= \frac{A^T}{det (A)}
\end{equation} 
where $A^T$ is a matrix composed of algebraic complements for the corresponding elements of the transposed matrix.

To determine $g^{-1}_{\mu\nu}$, we rewrite the matrix (\ref{mat}) in the following form.
\begin{equation}
g_{\mu\nu} =2B^2 \left(
\begin{array}{ccc}
(J_1\cos^2\theta_3+J_2\sin^2\theta_3)\sin^2\theta_2+J_3\cos^2\theta_2&
(J_2-J_1)\sin\theta_2\cos\theta_3\sin\theta_3&J_3\cos\theta_2\\
(J_2-J_1)\sin\theta_2\cos\theta_3\sin\theta_3&
J_1\sin^2\theta_3+J_2\cos^2\theta_3&0\\
J_3\cos\theta_2&0&J_3.
\end{array}
\right)
\label{mat1}
\end{equation}
First, we find the matrix of minors
\begin{equation}
 M_{ij}= \left(
\begin{array}{ccc}
*&*&*\\
*&*&*\\
*&*&*.
\end{array}
\right)
\end{equation}
Consider the first element of the matrix $g_{\mu\nu}$ (\ref{mat1}) and cross out the row and column containing this element
\begin{equation}
g' =\left(
\begin{array}{cc}
J_1\sin^2\theta_3+J_2\cos^2\theta_3&0\\
0&J_3.
\end{array}
\right)
\end{equation}
 This determinant $"$two by two$"$ is the minor of this element. Let's calculate:
$$
D=(J_1\sin^2\theta_3+J_2\cos^2\theta_3)\cdot J_3,
$$
write it into our matrix of minors:
$$
M_{11}=(J_1\sin^2\theta_3+J_2\cos^2\theta_3)\cdot J_3,
$$
Consider the second element of the matrix $g_{\mu\nu}$ (\ref{mat1}) and cross out the row and column containing this element
\begin{equation}
g'=\left(
\begin{array}{cc}
(J_2-J_1)\sin\theta_2\cos\theta_3\sin\theta_3&0\\
J_3\cos\theta_2&J_3.
\end{array}
\right)
\end{equation}
$$
M_{12}=(J_2-J_1)\sin\theta_2\cos\theta_3\sin\theta_3\cdot J_3,
$$
Consider the third element of the matrix $g_{\mu\nu}$ (\ref{mat1}) and cross out the row and column containing this element
\begin{equation}
g'=\left(
\begin{array}{cc}
(J_2-J_1)\sin\theta_2\cos\theta_3\sin\theta_3&
J_1\sin^2\theta_3+J_2\cos^2\theta_3\\
J_3\cos\theta_2&0.
\end{array}
\right)
\end{equation}
$$
M_{13}=-J_3\cos\theta_2\cdot (J_1\sin^2\theta_3+J_2\cos^2\theta_3),
$$
Consider the fourth element of the matrix $g_{\mu\nu}$ (\ref{mat1}) and cross out the row and column containing this element
\begin{equation}
g'=\left(
\begin{array}{cc}
(J_2-J_1)\sin\theta_2\cos\theta_3\sin\theta_3&J_3\cos\theta_2\\
0&J_3.
\end{array}
\right)\end{equation}
$$
M_{21}=(J_2-J_1)\sin\theta_2\cos\theta_3\sin\theta_3\cdot J_3,
$$
Consider the fifth element of the matrix $g_{\mu\nu}$ (\ref{mat1}) and cross out the row and column containing this element 
\begin{equation}
g'=\left(\begin{array}{cc}
(J_1\cos^2\theta_3+J_2\sin^2\theta_3)\sin^2\theta_2+J_3\cos^2\theta_2&J_3\cos\theta_2\\
J_3\cos\theta_2&J_3.
\end{array}
\right)
\end{equation}
$$
M_{22}=[(J_1\cos^2\theta_3+J_2\sin^2\theta_3)\sin^2\theta_2+J_3\cos^2\theta_2]\cdot J_3-J^2_3\cos^2\theta_2=
$$
$$
=(J_1\cos^2\theta_3+J_2\sin^2\theta_3)\sin^2\theta_2\cdot J_3+J^2_3\cos^2\theta_2-J^2_3\cos^2\theta_2=(J_1\cos^2\theta_3+J_2\sin^2\theta_3)\sin^2\theta_2\cdot J_3,
$$
Consider the sixth element of the matrix $g_{\mu\nu}$ (\ref{mat1}) and cross out the row and column containing this element
\begin{equation}
g'=\left(
\begin{array}{cc}
(J_1\cos^2\theta_3+J_2\sin^2\theta_3)\sin^2\theta_2+J_3\cos^2\theta_2&
(J_2-J_1)\sin\theta_2\cos\theta_3\sin\theta_3\\
J_3\cos\theta_2&0.
\end{array}
\right)
\end{equation}
$$
M_{23}=-J_3\cos\theta_2(J_2-J_1)\sin\theta_2\cos\theta_3\sin\theta_3
$$
Consider the seventh element of the matrix $g_{\mu\nu}$ (\ref{mat1}) and cross out the row and column containing this element
\begin{equation}
g'=\left(
\begin{array}{cc}
(J_2-J_1)\sin\theta_2\cos\theta_3\sin\theta_3&J_3\cos\theta_2\\
J_1\sin^2\theta_3+J_2\cos^2\theta_3&0.
\end{array}
\right)
\end{equation}
$$
M_{31}=-J_3\cos\theta_2\cdot (J_1\sin^2\theta_3+J_2\cos^2\theta_3)
$$
Consider the eighth element of the matrix $g_{\mu\nu}$ (\ref{mat1}) and cross out the row and column containing this element
\begin{equation}
g'=\left(
\begin{array}{cc}
(J_1\cos^2\theta_3+J_2\sin^2\theta_3)\sin^2\theta_2+J_3\cos^2\theta_2&J_3\cos\theta_2\\
(J_2-J_1)\sin\theta_2\cos\theta_3\sin\theta_3&0\\
\end{array}
\right)
\end{equation}
$$
M_{32}=-J_3\cos\theta_2\cdot (J_2-J_1)\sin\theta_2\cos\theta_3\sin\theta_3
$$
Consider the ninth element of the matrix $g_{\mu\nu}$ (\ref{mat1}) and cross out the row and column containing this element 
\begin{equation}
g'=\left(
\begin{array}{cc}
(J_1\cos^2\theta_3+J_2\sin^2\theta_3)\sin^2\theta_2+J_3\cos^2\theta_2&
(J_2-J_1)\sin\theta_2\cos\theta_3\sin\theta_3\\
(J_2-J_1)\sin\theta_2\cos\theta_3\sin\theta_3&
J_1\sin^2\theta_3+J_2\cos^2\theta_3.
\end{array}
\right)
\end{equation}
$$
M_{33}=[(J_1\cos^2\theta_3+J_2\sin^2\theta_3)\sin^2\theta_2+J_3\cos^2\theta_2]\cdot [J_1\sin^2\theta_3+J_2\cos^2\theta_3]-(J_2-J_1)^2\sin^2\theta_2\cos^2\theta_3\sin^2\theta_3
$$
$$ M_{ij}= 2B^2\cdot$$
\begin{equation}
\cdot
\begin{pmatrix}
(J_1\sin^2\theta_3+J_2\cos^2\theta_3)\cdot J_3 &  (J_2-J_1)\sin\theta_2 &-J_3\cos\theta_2\cdot \\ 
&  \cos\theta_3\sin\theta_3\cdot J_3 & (J_1\sin^2\theta_3+J_2\cos^2\theta_3)\\ \\
(J_2-J_1)\sin\theta_2\cdot&  (J_1\cos^2\theta_3+J_2\sin^2\theta_3)& -J_3\cos\theta_2(J_2-J_1)\sin\theta_2\cos\theta_3\sin\theta_3\\ 
\cos\theta_3\sin\theta_3 J_3 & \sin^2\theta_2\cdot J_3 & \\ \\
-J_3\cos\theta_2\cdot& -J_3\cos\theta_2\cdot (J_2-J_1) & [(J_1\cos^2\theta_3+J_2\sin^2\theta_3)\sin^2\theta_2+J_3\cos^2\theta_2]\cdot\\ 
(J_1\sin^2\theta_3+J_2\cos^2\theta_3)& \sin\theta_2\cos\theta_3\sin\theta_3 & [J_1\sin^2\theta_3+J_2\cos^2\theta_3]-\\
& & (J_2-J_1)^2\sin^2\theta_2\cos^2\theta_3\sin^2\theta_3.
\end{pmatrix}
\label{minor}
\end{equation}
In the matrix of minors (\ref{minor}), it is necessary to change signs strictly for the following elements: $M_{12}$, $M_{21}$, $M_{23}$, $M_{32}$.
$$ M_{ij}= 2B_3^2\cdot$$
\begin{equation}
\cdot
\begin{pmatrix}
(J_1\sin^2\theta_3+J_2\cos^2\theta_3)\cdot J_3 &  -(J_2-J_1)\sin\theta_2 &-J_3\cos\theta_2\cdot \\ 
&  \cos\theta_3\sin\theta_3\cdot J_3 & (J_1\sin^2\theta_3+J_2\cos^2\theta_3)\\ \\
-(J_2-J_1)\sin\theta_2\cdot&  (J_1\cos^2\theta_3+J_2\sin^2\theta_3)& J_3\cos\theta_2(J_2-J_1)\sin\theta_2\cos\theta_3\sin\theta_3\\ 
\cos\theta_3\sin\theta_3 J_3 & \sin^2\theta_2\cdot J_3 & \\ \\
-J_3\cos\theta_2\cdot& J_3\cos\theta_2\cdot (J_2-J_1) & [(J_1\cos^2\theta_3+J_2\sin^2\theta_3)\sin^2\theta_2+J_3\cos^2\theta_2]\cdot\\ 
(J_1\sin^2\theta_3+J_2\cos^2\theta_3)& \sin\theta_2\cos\theta_3\sin\theta_3 & [J_1\sin^2\theta_3+J_2\cos^2\theta_3]-\\
& & -(J_2-J_1)^2\sin^2\theta_2\cos^2\theta_3\sin^2\theta_3.
\end{pmatrix}
\label{minor1}
\end{equation}
(\ref{minor1}) matrix of algebraic complements of the corresponding elements.

Find the transposed matrix of algebraic additions $M^T_{ij}$.
$$ M^T_{ij}= 2B_3^2\cdot$$
\begin{equation}
\cdot
\begin{pmatrix}
(J_1\sin^2\theta_3+J_2\cos^2\theta_3)\cdot J_3 &  -(J_2-J_1)\sin\theta_2 &-J_3\cos\theta_2\cdot \\ 
&  \cos\theta_3\sin\theta_3\cdot J_3 & (J_1\sin^2\theta_3+J_2\cos^2\theta_3)\\ \\
-(J_2-J_1)\sin\theta_2\cdot&  (J_1\cos^2\theta_3+J_2\sin^2\theta_3)& J_3\cos\theta_2(J_2-J_1)\sin\theta_2\cos\theta_3\sin\theta_3\\ 
\cos\theta_3\sin\theta_3 J_3 & \sin^2\theta_2\cdot J_3 & \\ \\
-J_3\cos\theta_2\cdot& J_3\cos\theta_2\cdot (J_2-J_1) & [(J_1\cos^2\theta_3+J_2\sin^2\theta_3)\sin^2\theta_2+J_3\cos^2\theta_2]\cdot\\ 
(J_1\sin^2\theta_3+J_2\cos^2\theta_3)& \sin\theta_2\cos\theta_3\sin\theta_3 & [J_1\sin^2\theta_3+J_2\cos^2\theta_3]-\\
& & -(J_2-J_1)^2\sin^2\theta_2\cos^2\theta_3\sin^2\theta_3.
\end{pmatrix}
\label{minor2}
\end{equation}
(\ref{minor2}) transposed matrix of algebraic complements.
Get the inverse matrix
\begin{equation}
g^{-1}_{\mu\nu}=\frac{M^T_{\mu\nu}}{|G|}.
\end{equation}
The following should be noted here:
$$
g^{-1}_{55}=\sin^2\theta_2[J_1J_2\cos^4\theta_3+J_2\cos^2\theta_3(J_3\cot^2\theta_2+2J_1\sin^2\theta_3)
+J_1\sin^2\theta_3(J_3\cot^2\theta_2+J_2\sin^2\theta_3)].
\footnote{This expression is different from the one in \cite{greiner}.}$$

We choose generalized coordinates $q_\mu$
$$
q_1=a_{30},\hspace{0.3cm}q_2=a_{32}\hspace{0.3cm}q_3=\theta_1,\hspace{0.3cm}q_4=\theta_2, \hspace{0.3cm} q_5=\theta_3.
$$
$$
a_{30}=\beta_3\cos\eta,\hspace{0.3cm}a_{32}=\beta_3\frac{\sin\eta}{\sqrt{2}}.
$$
$$
G=64B^5_3\sin^2\theta_2 a_{32}^2\left[\left(3a^2_{30}+4a^2_{32}\right)^2-30a^2_{30}
a^2_{32}\right].
$$
$$
G^{1/2}=8B^{5/2}_3\sin\theta_2 \left[9a^4_{30}a^2_{32}-6a^2_{30}
a^4_{32}+16a^6_{32}\right]^{1/2}.
$$$$
g^{-1}_{11}=\frac{1}{B_3}
$$
$$
-\frac{\hbar^2}{2}\frac{1}{G^{1/2}}\frac{\partial}{\partial q_1}G^{1/2}g^{-1}_{11}\frac{\partial}{\partial q_1}=-\frac{\hbar^2}{2B_3}\frac{1}{G^{1/2}}\frac{\partial}{\partial a_{30}}G^{1/2}\frac{\partial}{\partial a_{30}}=
$$
$$
=-\frac{\hbar^2}{2B_3}\left\{\frac{6a_{30}\left(3a^2_{30}-
a^2_{32}\right)}{9a^4_{30}-6a^2_{30}
a^2_{32}+16a^4_{32}}\frac{\partial}{\partial a_{30}}+\frac{\partial^2}{\partial a^2_{30}}\right\}.
$$
$$
g^{-1}_{22}=\frac{1}{2B_3}
$$
$$
-\frac{\hbar^2}{2}\frac{1}{G^{1/2}}\frac{\partial}{\partial q_2}G^{1/2}g^{-1}_{22}\frac{\partial}{\partial q_2}=-\frac{\hbar^2}{4B_3}\frac{1}{G^{1/2}}\frac{\partial}{\partial a_{32}}G^{1/2}\frac{\partial}{\partial a_{32}}=
$$
$$
=-\frac{\hbar^2}{2B_3}\left\{\frac{3(3a^4_{30}-4a^2_{30}
a^2_{32}+16a^4_{32})}{2a_{32}(9a^4_{30}-6a^2_{30}
a^2_{32}+16a^4_{32})}\frac{\partial}{\partial a_{32}}+\frac{\partial^2}{2\partial a^2_{32}}\right\}.
$$
\begin{equation}
T_{vib}=-\frac{\hbar^2}{2B_3}\left[\frac{6a_{30}\left(3a^2_{30}-
a^2_{32}\right)}{9a^4_{30}-6a^2_{30}
a^2_{32}+16a^4_{32}}\frac{\partial}{\partial a_{30}}+\frac{\partial^2}{\partial a^2_{30}}+\frac{3(3a^4_{30}-4a^2_{30}
a^2_{32}+16a^4_{32})}{2a_{32}(9a^4_{30}-6a^2_{30}
a^2_{32}+16a^4_{32})}\frac{\partial}{\partial a_{32}}+\frac{\partial^2}{2\partial a^2_{32}}\right].
\label{vib}
\end{equation}
\begin{center} 
{\bf \uppercase{III. Transition to coordinates $\beta_3$ and $\eta$.}}
 \end{center}
Take into account in (\ref{vib}) expression (\ref{a20}), i.e.
$$
a_{30}=\beta_3\cos\eta,\hspace{0.5cm}a_{32}=\beta_3\frac{\sin\eta}{\sqrt{2}}.
$$
We rewrite following multipliers to $\beta_3$ and $\eta$ in (\ref{vib})  
$$
\frac{6a_{30}\left(3a^2_{30}-
a^2_{32}\right)}{9a^4_{30}-6a^2_{30}
a^2_{32}+16a^4_{32}}=\frac{3\cos\eta\left(5+7\cos2\eta\right)}{\beta_3(5+5\cos2\eta+8\cos^22\eta)}.
$$
$$
\frac{3(3a^4_{30}-4a^2_{30}
a^2_{32}+16a^4_{32})}{2a_{32}(9a^4_{30}-6a^2_{30}
a^2_{32}+16a^4_{32})}==\frac{3(5-2\cos2\eta+9\cos^22\eta)}{2\sqrt{2}\beta_3\sin\eta(5+5\cos2\eta+8\cos^22\eta)}.
$$
Then
$$
T_{vib}=-\frac{\hbar^2}{2B_3}\left[\frac{3\cos\eta\left(5+7\cos2\eta\right)}{\beta_3(5+5\cos2\eta+8\cos^22\eta)}\frac{\partial}{\partial a_{30}}+\frac{\partial^2}{\partial a^2_{30}}+\right.
$$
\begin{equation}
\left.+\frac{3(5-2\cos2\eta+9\cos^22\eta)}{2\sqrt{2}\beta_3\sin\eta(5+5\cos2\eta+8\cos^22\eta)}\frac{\partial}{\partial a_{32}}+\frac{\partial^2}{2\partial a^2_{32}}\right].
\label{vib}
\end{equation}
Now we determine  $\partial/\partial a_{30}$, $\partial^2/\partial a^2_{30}$, $\partial/\partial a_{32}$ and $\partial^2/\partial a^2_{32}$ in (\ref{vib}).
$$
\beta_3^2=a^2_{30}+2a^2_{32},\hspace{0.3cm}\beta_3=\sqrt{a^2_{30}+2a^2_{32}}.
$$
$$
\frac{\partial\beta}{\partial a_{30}}=\cos\eta.
$$
$$
\frac{\partial\beta}{\partial a_{32}}=\sqrt{2}\sin\eta.
$$
$$
\frac{a_{32}}{a_{30}}=\frac{1}{\sqrt{2}}\frac{\sin\eta}{\cos\eta}=\frac{\tan\eta}{\sqrt{2}}
$$
$$
\tan\eta=\sqrt{2}\frac{a_{32}}{a_{30}},\hspace{0.3cm}\eta=\arctan{\sqrt{2}\frac{a_{32}}{a_{30}}}
$$
$$
\frac{\partial\eta}{\partial a_{30}}=-\frac{\sin\eta}{\beta_3}
$$
$$
\frac{\partial\eta}{\partial a_{32}}=\frac{\sqrt{2}\cos\eta}{\beta_3}
$$
$$
\frac{\partial}{\partial a_{30}}=
\cos\eta\frac{\partial}{\partial \beta_3}-\frac{\sin\eta}{\beta_3}\frac{\partial}{\partial \eta}
$$
$$
\frac{\partial}{\partial a_{32}}=
\sqrt{2}\sin\eta\frac{\partial}{\partial \beta_3}+\frac{\sqrt{2}\cos\eta}{\beta_3}\frac{\partial}{\partial \eta}
$$
$$
\frac{\partial^2}{\partial a^2_{30}}=
\frac{\partial}{\partial a_{30}}\left(\frac{\partial}{\partial a_{30}}\right)=
\frac{\partial}{\partial \beta_3}\left(\frac{\partial}{\partial a_{30}}\right)\frac{\partial\beta_3}{\partial a_{30}}+\frac{\partial}{\partial \eta}\left(\frac{\partial}{\partial a_{30}}\right)\frac{\partial\eta}{\partial a_{30}}=
$$
$$
=\cos^{2}\eta\frac{\partial^2}{\partial \beta_3^{2}}-\frac{2\sin\eta\cos\eta}{\beta_3}\frac{\partial^2}{\partial \eta\partial \beta_3}+
\frac{2\sin\eta\cos\eta}{\beta^2}\frac{\partial}{\partial \eta}+\frac{\sin^2\eta}{\beta_3}\frac{\partial}{\partial \beta_3}+\frac{\sin^2\eta}{\beta_3^2}\frac{\partial^2}{\partial \eta^2}
$$
$$
\frac{\partial^2}{\partial a^2_{32}}=\frac{\partial}{\partial a_{32}}\left(\frac{\partial}{\partial a_{32}}\right)=\frac{\partial}{\partial \beta_3}\left(\frac{\partial}{\partial a_{32}}\right)\frac{\partial\beta_3}{\partial a_{32}}+\frac{\partial}{\partial \eta}\left(\frac{\partial}{\partial a_{32}}\right)\frac{\partial\eta}{\partial a_{32}}=
$$
$$
=2\sin^{2}\eta\frac{\partial^2}{\partial \beta_3^{2}}+\frac{4\sin\eta\cos\eta}{\beta}\frac{\partial^2}{\partial \eta\partial \beta_3}-
\frac{4\sin\eta\cos\eta}{\beta^2}\frac{\partial}{\partial \eta}+\frac{2\cos^2\eta}{\beta_3}\frac{\partial}{\partial \beta_3}+\frac{2\cos^2\eta}{\beta_3^2}\frac{\partial^2}{\partial \eta^2}
$$
Thus
\begin{equation}
\hat{T}_{vib}=-\frac{\hbar^2}{2B_3}\left[\frac{\partial^2}{\partial \beta_3^{2}}+\frac{4}{\beta_3}\frac{\partial}{\partial \beta_3}+
\frac{1}{\beta_3^2}\frac{\partial^2}{\partial \eta^2}+\frac{24\cos^22\eta-6\cos2\eta}
{5+5\cos2\eta+8\cos^22\eta}\frac{\cos\eta}{\beta^2_3\sin\eta}\frac{\partial}{\partial \eta}\right]
\label{beta3}
\end{equation}

A complete numerical solution of the Schr\"{o}dinger equation with the operator (\ref{beta3}) has not yet been found, and therefore, in all models of the atomic nucleus, the study of collective excitations is carried out by introducing simplifying assumptions. To do this, we first rewrite the formula (\ref{beta3}) as follows:
\begin{equation}
\hat{T}_{\beta_3}=-\frac{\hbar^2}{2B_3}\left[\frac{1}{\beta_3^4}\frac{\partial}{\partial \beta_3}\left(\beta_3^4\frac{\partial}{\partial \beta_3}\right)+
\frac{1}{\beta_3^2}\frac{\partial^2}{\partial \eta^2}+\frac{24\cos^22\eta-6\cos2\eta}
{5+5\cos2\eta+8\cos^22\eta}\frac{\cos\eta}{\beta^2_3\sin\eta}\frac{\partial}{\partial \eta}\right],
\label{beta31}
\end{equation}
 with volume element (\ref{vol})
\begin{equation}
d\tau= a_{32}\sqrt{\left(3a^2_{30}+4a^2_{32}\right)^2-30a^2_{30}
a^2_{32}}d a_{20}d a_{32}d\theta_1\sin\theta_2d\theta_2d\theta_3
\end{equation}
or
\begin{equation}
d\tau=\beta^4_3d\beta\sin\eta\sqrt{10+10\cos2\eta+16\cos^22\eta}d\eta d\theta_1\sin\theta_2d\theta_2d\theta_3.
\end{equation}
 The expression (\ref{beta31}) describes even-even nuclei with free non-axiality.
 
Further, we assume that the variable $\eta$ takes a fixed value, i.e. $\eta$=$\eta_{eff}$. Then 
\begin{equation}
\hat{T}_{\beta_3}=-\frac{\hbar^2}{2B_3}\frac{1}{\beta_3^3}\frac{\partial}{\partial \beta_3}\left(\beta_3^3\frac{\partial}{\partial \beta_3}\right),
\label{beta32}
\end{equation}
 with volume element
\begin{equation}
d\tau=\beta_3^3d\beta_3 d\theta_1\sin\theta_2d\theta_2d\theta_3.
\end{equation}
This expression describes even-even nuclei with effective non-axiality.

Further assumption is considered axially symmetric nuclei, i.e. $\eta$=0. In this case $a_{32}$=$a_{3-2}$=0 \cite{den,den1,min,min1,nad,sun}. Then
\begin{equation}
\hat{T}_{\beta_3}=-\frac{\hbar^2}{2B_3}\frac{1}{\beta_3^2}\frac{\partial}{\partial \beta_3}\left(\beta_3^2\frac{\partial}{\partial \beta_3}\right),
\label{beta33}
\end{equation}
 with volume element
\begin{equation}
d\tau=\beta_3^2d\beta_3 d\theta_1\sin\theta_2d\theta_2d\theta_3.
\end{equation}
It is known that the total energy of quadrupole deformation has the form:
\begin{equation}
T_{\beta_2}=\frac{B_2}{2}(\dot{a}^2_{20}+2\dot{a}^2_{22})+\sum^3_{\lambda=1}\frac{\hbar^2\hat{I}^2_\lambda}{J^{(2)}_\lambda},
\label{oct}
\end{equation}
here $J^{(2)}_\lambda$ of the projections of the total moment of inertia for the quadrupole deformation:
\begin{equation}
J^{(2)}_1=4B_2\left(3a^2_{20}+2a^2_{22}+2\sqrt{6}a_{20}
a_{22}\right)
\label{j11}
\end{equation}
\begin{equation}
J^{(2)}_2=4B_2\left(3a^2_{20}+2a^2_{22}-2\sqrt{6}a_{20}
a_{22}\right)
\label{j21}
\end{equation}
\begin{equation}
J^{(2)}_3=8B_2a_{22}^2
\label{j31}
\end{equation}
\begin{equation}
 a'_{20}=\beta_2\cos\gamma,\hspace{2cm}
a'_{22}=a'_{3,-2}=\frac{\beta_2\sin\gamma}{\sqrt{2}},
\label{a30}
\end{equation}
Thus, we get the expression for the classical kinetic energy (\ref{k1}) in curvilinear coordinates: 
\begin{equation}
\hat{T}_{\beta_2}=-\frac{\hbar^2}{2B_2}\left[\frac{4}{\beta_2}\frac{\partial}{\partial \beta_2}+\frac{\partial^2}{\partial \beta_2^2}+3\cot3\gamma\frac{1}{\beta_2^2}\frac{\partial}{\partial\gamma}+\frac{1}{\beta_2^2}\frac{\partial^2}{\partial \gamma^{2}}\right],
\label{beta2}
\end{equation}
 with volume element
\begin{equation}
d\tau=\beta_2^4d\beta_2\sin 3\gamma d\gamma d\theta_1\sin\theta_2d\theta_2d\theta_3.
\end{equation}

The resulting expression (\ref{beta3}) is compared with a similar expression for quadrupole oscillations of the nuclear surface (\ref{beta2}). It can be seen that the last multipliers in (\ref{beta3}) and (\ref{beta2}) do not match. Although the dynamic variables $a_{20}$, $a_{22}$ and $a_{30}$, $a_{32}$ have similar expressions (\ref{a20}) and (\ref{a30}), but the form of expression of the projection of their moments of inertia is different, i.e. formulas (\ref{j12})--(\ref{j13}) and (\ref{j11})--(\ref{j31}).

Thus in this work, the quantization of the kinetic energy associated with the quadrupole (octupole) oscillations of the surface of the nucleus in curvilinear coordinates was carried out. The resulting form of the Hamiltonian differs from the previously obtained Hamiltonian for quadrupole oscillations only by factors in front of the derivatives $\partial/\partial\gamma$ and $\partial/\partial\eta$, because the expressions for the projections of the moments of inertia of the nucleus in the cases of quadrupole and octupole oscillations are different. In addition, an explicit form of the kinetic energy of the Hamiltonian of even-even nuclei with arbitrary and effective non-axiality is obtained, as well as an explicit form of the kinetic energy of the Hamiltonian for axially symmetric even-even nuclei.

\end{document}